\documentclass{mn2e}
\usepackage{epsfig}
\usepackage{times}
\begin{document}

\title[SGR 1806-20] 
{Structure in the radio counterpart to the 2004 Dec 27 giant flare from SGR 1806-20}
\author[Fender et al.]  
{R. P. Fender$^1$\thanks{email:rpf@phys.soton.ac.uk},
T.W.B. Muxlow$^2$, M.A. Garrett$^3$, C. Kouveliotou$^4$,
B.M. Gaensler$^5$,  \newauthor S.T. Garrington$^2$, Z. Paragi$^3$,
V. Tudose$^{6,7}$, J.C.A. Miller-Jones$^6$, R.E. Spencer$^2$,
 \newauthor R.A.M. Wijers$^{6}$,  G.B. Taylor$^{8,9,10}$\\
$^1$ School of Physics and Astronomy, University of Southampton,
 Highfield, Southampton, SO17 1BJ, UK\\
$^2$ University of Manchester, 
Jodrell Bank Observatory, Cheshire, SK11 9DL, UK\\
$^3$ Joint Institute for VLBI in Europe, Postbus 2, NL-7990 AA Dwingeloo, The Netherlands\\
$^4$ NASA / Marshall Space Flight Center, NSSTC, XD-12, 320 Sparkman Drive, Huntsville, AL 35805, USA\\
$^5$ Harvard-Smithsonian Center for Astrophysics, 60 Garden Street, Cambridge, MA 02138, USA\\
$^6$ Astronomical Institute `Anton Pannekoek', University of
Amsterdam, Kruislaan 403, 1098 SJ Amsterdam, The Netherlands\\
$^7$ Astronomical Institute of the Romanian Academy, Cutitul de Argint 5, RO-040557
Bucharest, Romania\\
$^8$ Kavli Institute of Particle Astrophysics and Cosmology, Menlo Park, CA 94025 USA\\
$^9$ National Radio Astronomy Observatory, Socorro, NM 87801, USA\\
$^{10}$Department of Physics and Astronomy, University of New Mexico, Albuquerque, NM 87131, USA \\}
\maketitle

\begin{abstract}
On Dec 27, 2004, the magnetar SGR 1806-20 underwent an enormous
outburst resulting in the formation of an expanding, moving, and
fading radio source. We report observations of this radio source with
the Multi-Element Radio-Linked Interferometer Network (MERLIN) and the
Very Long Baseline Array (VLBA). The observations confirm the
elongation and expansion already reported based on observations at
lower angular resolutions, but suggest that at early epochs the
structure is not consistent with the very simplest models such as a
smooth flux distribution.  In particular there appears to be
significant structure on small angular scales, with $\sim 10$\% of the
radio flux arising on angular scales $\leq 100$ milliarcsec.  This structure
may correspond to localised sites of particle acceleration during the
early phases of expansion and interaction with the ambient medium.
\end{abstract}

\section{Introduction}

On Dec 27, 2004, the most energetic explosion witnessed by humans
within our galaxy for over 400 years was detected from the soft
gamma-ray repeater SGR 1806-20 (e.g. Borkowski et al. 2004; Palmer et
al. 2004; Hurley et al. 2005).  Shortly after the outburst an
expanding radio source was detected associated with SGR 1806-20
(Cameron et al. 2005; Gaensler et al. 2005). This radio emission
traces the ejection of mass from the surface of the neutron star, and
its interaction with the ambient medium (Gelfand et al. 2005; Taylor
et al. 2005; Granot et al. 2005) Measuring the geometry and temporal
evolution of this ejected matter is of key importance for our
understanding of the origins of this enormous outburst, and its impact
upon its immediate environment.

SGR 1806-20 is believed to be a magnetar, an isolated non-accreting
neutron star with a magnetic field $\geq 10^{14}$ G (Kouveliotou et
al. 1998).  While several scenarios discussed for the origin of the
radio emission from SGR 1806-20 predict possibly edge-brightened
emission (Gaensler et al. 2005; Gelfand et al. 2005; Granot 2005) or
other spatial structure, especially at early times, the relatively low
resolution radio data published to date have not allowed a test of
these models. As a result, Gaussian models have been fit to the data
sets in order to quantify the motion and expansion of the source,
despite the link between the fit parameters and physical conditions in
the radio source remaining uncertain. In this paper we present the
highest-resolution radio images of the ejecta between nine and
fifty-six days after the outburst, obtained with the US Very Long
Baseline Array (VLBA) and the Multi-Element Radio-Linked
Interferometer Network (MERLIN) in the UK. These are the first, and
possibly only, data sets which are able to probe on angular scales
$\la 100$ mas and test for substructure in the evolving radio source.

\section{Observations}

A log of the six epochs of observation under discussion is presented
in table 1. Note that, with the exception of epoch M4, for all data
sets phase-referencing failed, due to resolved calibrators.
As a result, self-calibration was used
exclusively, removing any possibility of measuring the absolute
position / displacement of the radio source. Therefore we cannot test
the evidence for proper motion, and not just expansion, of the source
presented in Taylor et al. (2005). Fig 1 places these observations in
the context of the light curve of the radio source. The data were
calibrated using AIPS (Diamond 1995) while the images presented here were created in
MIRIAD (Sault, Teuben \& Wright 1995). The images themselves are based
upon discrete Fourier transforms (DFTs), which are better at avoiding
aliasing problems than fast Fourier transforms (FFTs). In practice the
difference is not great; in the region within 0.5 arcsec of SGR
1806-20 the difference map between the FFT and DFT images has an
r.m.s. of 35$\mu$Jy (typically 10--20\% of the total r.m.s. noise in
an image).  Nevertheless, there is structure in this difference map
and we default to the DFT images for all data sets in this paper. All
images presented in this paper have also been CLEANed, with a typical
gain of 0.05 over 500 iterations. MERLIN and VLBA images are presented
in Figs 2 and 3 respectively.

\begin{table}
\caption{Log of observations}
\begin{tabular}{rcccl}
Run & & Frequency & Midpoint (UT) / duration \\
\hline
V1 & VLBA   & 1.4 GHz & 2005-01-05-16:30 / 5.5 hr\\
M1 & MERLIN & 6.0 GHz & 2005-01-06-10:30 / 4.5 hr\\
V2 & VLBA   & 1.4 GHz & 2005-01-06-14:00 / 2.5 hr\\
M2 & MERLIN & 6.0 GHz & 2005-01-07-11:30 / 4.5 hr\\
M3 & MERLIN & 6.0 GHz & 2005-01-09-12:30 / 4.5 hr\\
M4 & MERLIN & 1.7 GHz & 2005-02-21-08:15 / 5.5 hr\\
\hline
\end{tabular}
\end{table}

\begin{figure}
\centerline{\epsfig{file=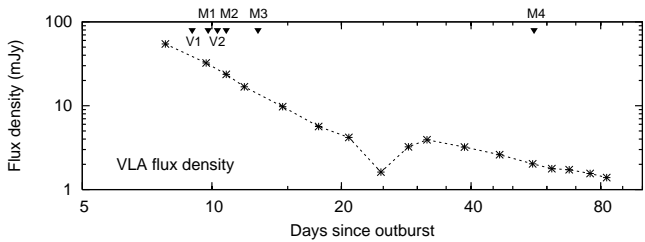, angle=270, width=9cm}}
\caption{VLA 8 GHz light curve of the radio counterpart of SGR 1806-20
(Taylor et al. 2005), with the epochs of MERLIN and VLBA
observations indicated. }
\end{figure}

\begin{table*}
\begin{tabular}{cccccccccc}
 & &  & \multicolumn{4}{c}{Image-plane Gaussian} & \multicolumn{3}{c}{{\em uv}-plane Gaussian}\\ 
 Epoch & map peak &  $F_{\rm comp}$ & peak & integrated & size & p.a. & integrated & size & p.a. \\
       & (mJy bm$^{-1}$) & (mJy) & (mJy bm$^{-1}$) & (mJy) & (mas$\times$mas) & ($^{\circ}$) & (mJy) & (mas$\times$mas) & ($^{\circ}$) \\ 
\hline\\
  M1 & $12.1 \pm 0.3$ & $30 \pm 2$ (6.1 GHz / A) & $10.5 \pm 0.4$ & $29.4 \pm 0.4$ & $234 \times 54$ & -16.6 & $26.9 \pm 0.6$ & $76 \times 67$ & $8.6 \pm 175$ \\
  M2 & $8.9 \pm 0.3$ & $21 \pm 1$ (6.1 GHz / A) & $8.5 \pm 0.4$ & $14.9 \pm 0.4$ & $147 \times 39$ & -20.3 & \multicolumn{3}{c}{Fit failed} \\
  M3 & $4.9 \pm 0.3$ & $21 \pm 1$ (6.1 GHz / A) & $4.0 \pm 0.2$ & $11.1 \pm 0.2$ & $234 \times 50$ & -23.5 & $9.8 \pm 0.5$ & $65 \times 28$ & $89.3 \pm 70$ \\
\hline\\
  M4 & $1.2 \pm 0.1$ & $6 \pm 1$ (1.4 GHz / V) & $1.1 \pm 0.1$ & $3.3 \pm 0.1$ & $680 \times 263$ & -36.8 & $3.5 \pm 0.4$ & $708 \times 215$ & $-31 \pm 9$ \\
\hline\\
  V1 & $6.7 \pm 0.2$ & $117 \pm 5$ (1.4 GHz / A) & $5.7 \pm 0.4$ & $63.4 \pm 0.4$ & $106 \times 79$ & -49.3 & $59.3 \pm 0.6$ & $92 \times 73$ & $-64.9 \pm 2.8$ \\
  V2 & $4.5 \pm 0.2$ & $91 \pm 2$ (1.4 GHz / A) & $3.1 \pm 0.3$ & $54.7 \pm 0.3$ & $206 \times 72$ & -43.1 & $46.6 \pm 1.5$ & $198 \times 72$ & $-38.8 \pm 3.7$ \\
\hline\\
\end{tabular}
\caption{Fits to the MERLIN and VLBA observations. We list the peak
flux as measured from the cleaned images, plus image-plane and {\em
uv}-plane single-Gaussian fits. In addition $F_{\rm comp}$ gives the
nearest (in time and frequency) comparison measurement from ATCA (A)
or VLA (V); data from Gaensler et al. (2005) except for M4 (Gelfand et
al. in prep).  For the MERLIN data the fits were unconstrained, and in
all cases (except the {\em uv}-plane fit for M2) converged. In order
to get the fits to converge for the VLBA datasets, it was necessary to
constrain the initial fit parameters; for the image-plane fits we
specified a small region for the fit, and it was found that the fitted
parameters were a weak function of the dimensions of this region. For
the {\em uv}-plane fits we used the image-plane fits as initial
guesses for the fit parameters. For both MERLIN and VLBA the
image-plane fits the fitted errors on the dimensions and positions
angles of the Gaussians are typically severe underestimates of the
total uncertainties, and are not listed here.  For the image-plane,
the positive discrepancy between the map peak and the peak of the
fitted Gaussian implies excess unresolved emission at this point.}
\end{table*}

\begin{figure*}
\centerline{\epsfig{file=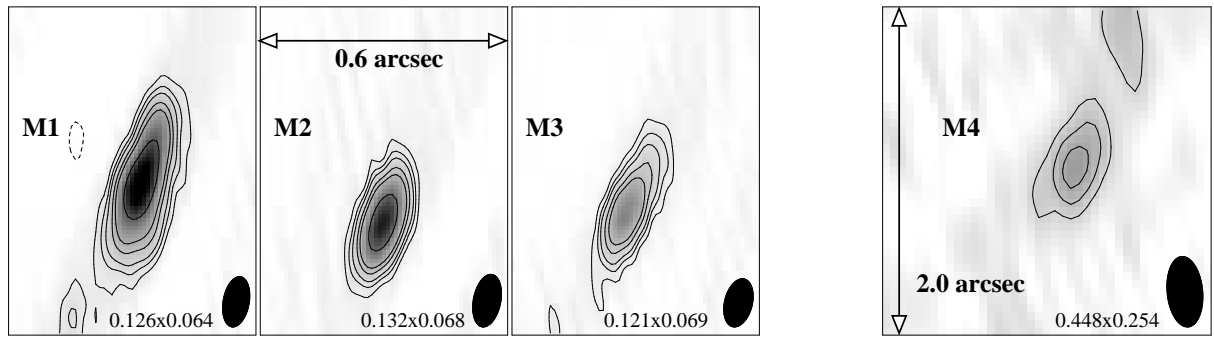, angle=0, width=16cm}}
\caption{Images of the radio source associated with the 2004 Dec 27
giant outburst of SGR 1806-20, obtained with MERLIN. The source is
clearly resolved at all epochs; however poor {\em uv} coverage and an
approximate alignment of the synthesised beam (indicated by solid
ellipses plus dimensions in arcsec) with the major axis of the radio
source render quantitative analysis problematic. The images were
computed using discrete fourier transforms; contour levels are
(-3,3,4.5,6,9,12,24,48,96) $\times$ the r.m.s. noise in each image
(0.27, 0.24, 0.22 and 0.17 mJy bm$^{-1}$ for epochs M1,M2, M3 and M4
respectively).  The greyscale is the same for epochs M1-M3, to
illustrate the fading of the source; for epoch M4 the grey scale range
is scaled down by a factor 0.3.}
\end{figure*}

Note that throughout this paper we utilise and discuss Gaussian models
to the radio flux distribution. This is because they are a simple way
to parametrize a barely-resolved source and were therefore used to fit
the published VLA data sets (Gaensler et al. 2005; Taylor et
al. 2005), against which we compare our data. Note that neither here
nor in the previously published papers are we claiming for theoretical
reasons that the flux distribution should be Gaussian in form.

\subsection{MERLIN}

As a southern object SGR 1806-20 is a very awkward target for MERLIN,
and as a result the data reduction has considerably more associated
difficulties than for a standard northern source. These problems were
exacerbated by the failure of the Cambridge telescope during runs M1,
M2 and M3 in 2005 January, further reducing the quality of the {\em
u-v} coverage, and by a resolved calibrator. 


The flux scales for epochs M1-M3 were calibrated with 3C286 and
0552+398. 3C286 is around 5\% resolved on the shortest MERLIN spacings
whilst 0552+398 is unresolved. The phase calibrator was
1808+209. Initial imaging of runs M1--3 appears to reveal significant
deviations from the Gaussian model employed to fit the Very Large
Array (VLA) data (e.g. Gaensler et al. 2005; Taylor et al. 2005),
which have considerably poorer angular resolution. However, there are
many caveats to this conclusion: the calibrator was resolved, and the
resultant images were found to be sensitive to the starting model.
Therefore all apparent structure in the images in Fig 2 must be
treated with caution.  The overall elongation of the resolved radio
structure in a north-westerly direction is not in doubt. However, note
that for all MERLIN epochs it appears that there is a systematic
rotation of the true source extension position angle back towards
north from northwest, by approximately the same amount at each
epoch. This effect is confirmed in simulations, which indicate that
the source extent and, possibly, structure, may be preserved even with
such poor {\em u-v} coverage, but that the position angle is less
trustworthy.

In Fig 2 we present DFT images of the three 6.0 GHz MERLIN epochs from
2005 January. Results of Gaussian fits in the image plane to these
maps are presented in table 2, as are measurements of the peak flux
density (per beam) in each map. It is clear from the discrepancy
between the measured and fitted peaks that these images cannot be fit
by simple Gaussians. The fits to the MERLIN datasets presented in
table 2 were unconstrained in any parameter and result in good fitting
of the extended emission but underfit the core; as a result the
residual images all show a point source with a flux density
approximately equal to the difference between the measured peaks in
table 2. Alternatively, fitting of Gaussians with the same size as
those fit to the VLA {\em uv}-plane data (Taylor et al. 2005) results
in good subtraction of the core emission, but residual extended
emission at all position angles. What remains unclear is to what
extent these discrepancies are due to real source structure, or have
an origin in artefacts.


The data reduction associated with run M4 was rather smoother, with a
well-behaved calibrator and Cambridge antenna available.  The flux
scale was set by observations of the point source OQ208 and resolved
flux density calibrator 3C286; the phase calibrator used was 1808-209.
A DFT image of this epoch is presented in Fig 2. Note that there
appears to be significant extended emission well beyond the central
smooth component; given the aformentioned uncertainties in mapping and
modelling the MERLIN data, we do not currently consider this to be
real. Fitting a single Gaussian to the main radio component (details
in table 2) does not result in significant undersubtraction of the
core as it did in the earlier, higher-frequency MERLIN epochs M1--3.
This hints at a diffusion of the compact radio structure which was
probably present at earlier epochs.

At the lower frequency of 1.4 GHz, run M4 does not probe the radio
structure at a much finer angular resolution than the early VLA
observations; however since the VLA is moving to more compact
configurations throughout 2005 the data remain valuable. In any case,
the source has clearly grown significantly in size between epochs M1-3
and M4, unless its angular size is a strong function of
wavelength. However, the overall optically thin sychrotron spectrum
across this frequency range at all epochs essentially rules out this
possibility.

\subsection{VLBA}

SGR 1806-20 was observed at 1.4 GHz with the NRAO Very Long Baseline
Array (VLBA), including the Green Bank Telescope and a single dish of
the VLA on 2005 January 5 and 6 (see Table 1 for exact times of
observations). The data were processed in AIPS following standard
amplitude calibration and fringe-fitting procedures (e.g. Diamond
1995). The amplitude calibration of the VLBA is typically better than
10\%. The target was phase-referenced to the nearby source
J1811-2055. However this source was found to be heavily
resolved. Instead, we applied delay and rate referencing to
J1825-1718, another calibrator that was regularly observed every 10-20
minutes. Self-calibration was carried out using Difmap (Shepherd
1997). Since a point source starting model was not applicable to our
data, and various extended models resulted in somewhat different final
images, we decided to use the MERLIN imaging result as a starting
model for phase calibration. Further VLBA observations were performed
on 2005 Jan 10 and Feb 28, but did not produce useful data (Jan 10 was
very weak and Feb 28 a non-detection) and are not discussed further
here.

DFT images are presented in Fig 3, and table 2 presents peak flux
measurements and Gaussian fits to the data.  It is immediately
apparent from Fig 3 that the radio sources do not look like simple
smooth flux distributions, but are highly elongated in a
north-westerly direction and display substructure.  As with the MERLIN
images, however, we have to carefully examine how much of any apparent
structure is real. An additional complication is that the combination
of high angular resolution, relatively low frequency and large
distance ($\sim 15$ kpc; Corbel \& Eikenberry 2004; McClure-Griffiths
\& Gaensler 2005) in the galactic plane makes it likely that some
interstellar scattering is affecting the images. At the position and
distance of SGR 1806-20 (galactic coordinates $l=10.0$,$b=-0.4$) the
Cordes \& Lazio (2002) electron density model predicts an angular
broadening at 1.4 GHz of 98$^{+14}_{-18}$ mas, although this may be an
overestimate given the evidence for structure on comparable or smaller
scales.


\section{Analysis}

It is clear from the MERLIN and VLBA observations that we have a
clearly resolved radio source oriented in an approximately
north-west direction. In the following we discuss further evidence for
the small-scale structure directly from the measured visibilities, and
compare our results with those from other telescopes.

\begin{figure}
\centerline{\epsfig{file=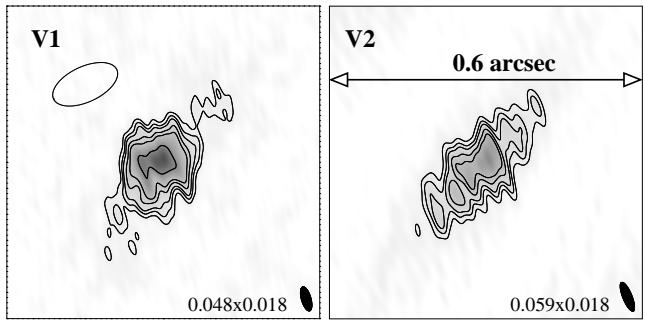, angle=0, width=8.5cm}}
\caption{VLBA images at 1.4 GHz from epochs V1 and V2. The contour
levels are (-3,3,4.5,6,9,12,24,48,96) $\times$ the r.m.s. noise in
each image (0.19 and 0.22 mJy  bm$^{-1}$ for epochs V1 and V2 respectively).  The
greyscale range is the same in each epoch to illustrate the fading of
the source. Solid ellipses are the synthesised beams (with dimensions in arcsec); the open ellipse
at epoch V1 is the Gaussian fit to the VLA data at approximately the
same epoch (Taylor et al. 2005).}
\end{figure}

\subsection{Reality of the small-scale structure}

All of the early images with MERLIN and the VLBA (i.e. those obtained
in 2005 January) show some evidence for substructure on small angular
scales in the SGR 1806-20 radio source.  Yet, as discussed above, it
is not clear at what level we can trust these finer
features. Inspection of the variation of visibility amplitude with
projected baseline length (expressed in units of wavelengths) can help
us establish the existence of this substructure.

The VLA observations of SGR 1806-20 have a longest projected baseline
length of $\sim 1000$ k$\lambda$, corresponding to an angular scale of
200 milli-arcsec. The MERLIN and VLBA data are able to probe angular
scales four and ten times smaller, respectively, as a result of
correspondingly longer projected baselines. Fig 4 illustrates the
additional information which may be obtained about the radio source by
utilising the longer MERLIN and VLBA baselines, for data around 2005
Jan 5/6.  The top panel illustrates the expected variation of
amplitude with baseline for circular Gaussian and circular thin shell
models (the former having been used to date to fit the
lower-resolution radio data).  It is clear that the circular shell
model predicts significantly more flux on the longer baselines,
reflecting the concentrations of flux on small angular scales, as
indeed would any model with substucture (once more we are simply
attempting to quantify the reality and degree of substructure, rather
than apply a theoretically-motivated model).  In fact the Gaussian
model predicts no detectable flux on projected baselines longer than
about 3500 k$\lambda$.  The lower three panels compare the measured
visibility amplitudes as a function of projected baseline for the VLA,
MERLIN and VLBA data sets of 2005 January 5/6 (all the data were taken
within 24 hr).  Both MERLIN and VLBA data sets show significant
deviations from the assumption of zero signal in the interval
3000--4000 k$\lambda$ (the expectation value for zero signal is based
upon the measured variance of the signal within each bin, calculated
in the MIRIAD routine UVAMP).  This is a strong indication that some
of the measured flux is from unresolved regions with a high surface
brightness.

\begin{figure}
\centerline{\epsfig{file=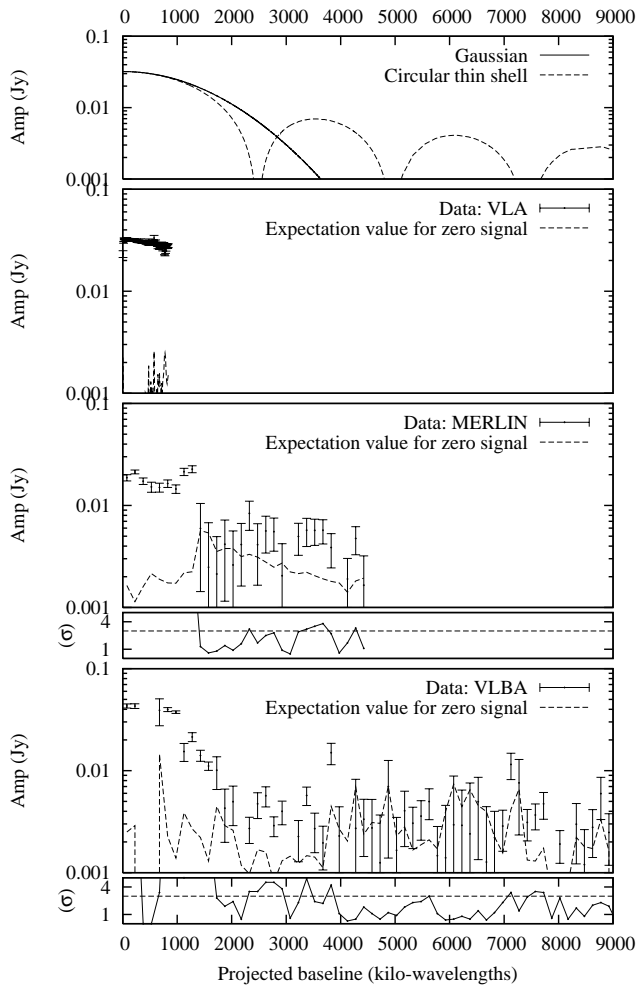, angle=0, width=9.5cm}}
\caption{Visibility amplitude as a function of projected baseline
distance obtained within 24hr of each other on 2005 Jan 5/6.  The top
panel plots the curves expected for circular Gaussian and circular
thin ring models, fit to the amplitude of the VLA measurement.  For
each of the three telescopes' data sets, the measured visibility
amplitude and the expectation value for zero signal are plotted.  The
VLA data are clearly very high significance detections and are
consistent with either model. However, only the MERLIN and VLBA data
are able to test the model on projected baselines at which the
differences become apparent, i.e. greater than about 2000 k$\lambda$
(corresponding to an angular resolution of about 100 mas). For these
data sets, the significance of the detections ($\sigma$) is also
plotted, with the 3$\sigma$ level indicated by the dashed
line. Calculation of the expectation value and significance of each
detection is based upon the measured variance within each bin, as
performed by the routine UVAMP in MIRIAD.  For both MERLIN and VLBA
data sets there is evidence for significant flux measured on longer
baselines which is not expected in the Gaussian model and indicates
compact structure at the $\la 10$\% level.}
\end{figure}

If we naively consider that the radio structure may be modelled as a
Gaussian plus compact source[s], then inspection of the visibility
curves in Fig 4 suggests that no more than a few mJy is associated
with these components. In other words, at this epoch $\geq 90$\% of
the radio flux seems to be associated with a diffuse component
well-modelled by a Gaussian, and $\leq 10$\% may be associated with
regions of higher surface brightness on angular scales $\leq 100$ mas.
This latter component could correspond to clumps of matter or to some
more regular structure such as a limb-brightened shell.  The
brightness temperatures associated with these compact regions would be
$\ga 10^4$K, which taken in isolation does not rule out thermal
processes (e.g. optically thin bremsstrahlung), although the overall
properties of the radio nebula, especially its spectrum and
polarisation (Gaensler et al. 2005; Taylor et al. 2005) are much more
consistent with optically thin synchrotron emission. Is this picture
of 90\% diffuse emission plus 10\% compact emission consistent with
the images ? The VLBA images
(Fig 3) clearly remind us that the diffuse emission component is not
evenly distributed, but rather more collimated, and does seem to be
combined with more compact structure.

\begin{figure}
\centerline{\epsfig{file=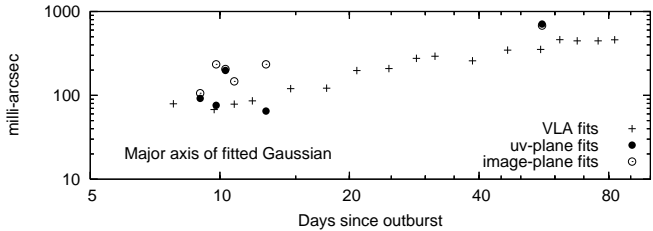, angle=270, width=9cm}}
\caption{Expansion of the SGR 1806-20 radio source: a comparison of
the major axes of the MERLIN and VLBA Gaussian fits (table 2) with
fits to the VLA data from Taylor et al. (2005).}
\end{figure}

\subsection{Comparison with other data}

As well as the image-plane fits, table 2 also presents {\em uv}-plane
fits to the MERLIN and VLBA datsets, allowing comparison with the fits
to the VLA data as presented in Taylor et al. (2005). Fig 5 compares
the fitted Gaussian major axes with those reported in Taylor et
al. (2005) from fits to VLA data; in addition to this a Gaussian fit
to contemporaneous VLA data is plotted alongside the V1 image in Fig
3.  While it seems clear from comparison of epoch M4 with the earlier
epochs that the source expansion is confirmed, it is very difficult to
quantify the rate of this expansion at early epochs. We further note
that a Gaussian fit to an image of epoch M4 which was restored with a
300 mas circular beam in AIPS resulted in a $300 \times 120$ mas
source (at a p.a. -44$^{\circ}$), which is closer to the values
reported in Taylor et al. (2005) but only serves to illustrate the
model- and approach-dependent nature of extracting numbers from the
images.  Inspection of epochs V1 and V2 suggests that a decrease in
the flux of the compact central component may be at least partially
responsible for the growth in the size of Gaussian fits.  Table 2 also
provides comparison with the nearest (in time and frequency) total
flux density measurements from Gaensler et al. (2005) and Gelfand et
al. (in prep). Apart from epoch M1, significant flux is being missed
by the larger arrays; in particular the VLBA. This is further evidence
for substructure as well as a lack of sensitivity to diffuse emission.

\section{Discussion}

In the context of stellar-mass compact objects, resolved
radio-emitting relativistic ejection events are typically associated
with periods of high rates of accretion and transitions in the state
of the accretion flow (e.g. Fender, Belloni \& Gallo 2004). The case
of SGR 1806-20 is clearly rather different: the object is probably not
accreting (Kouveliotou et al. 1998) and the apparent steady growth of
the radio source is rather different from the jets associated with
X-ray binaries, which typically have small ($\leq 10^{\circ}$) opening
angles (Miller-Jones, Fender \& Nakar 2005). Several aspects of this
behaviour may be associated with the probable lower bulk Lorentz
factor of this event, $\Gamma_{\rm SGR} \sim 1.4$ (Granot et al. 2005;
compare with $\Gamma \geq 2$ for X-ray binary jets). We can also
compare this event with the jet-like outflows from isolated neutron
stars such as the Crab (Hester et al. 2002), Vela (Pavlov et al. 2003)
and PSR~B1509-58 (DeLaney et al. 2005).  In these radio pulsars and
SGR~1806-20 it seems that the the 'escape velocity principle', in
which outflows have velocities comparable to the escape velocity at
their launch point (e.g. Livio 1999 and references therein) is
maintained, while it is is blatantly violated by the jets from the
accreting neutron stars Sco X-1 (Fomalont, Geldzahler \& Bradshaw
2001) and Circinus X-1 (Fender et al. 2004), which are much more
relativistic, suggesting the necessity of a disc in forming the most
relativistic flows.

Nevertheless, the radio source is still associated with the ejection
of matter in a preferred direction in space, and probably the
resultant in-situ particle acceleration. The observed synchrotron
emission is likely to have the highest surface brightness close to
regions of particle acceleration, and it may be these regions which
are producing the brightest regions of radio flux measured on the
longest baselines by MERLIN and the VLBA. Obvious sites for the
particle acceleration are at the leading edge of the ejecta (external
shocks), internal shocks distributed along the flow itself (easier to
achieve the more collimated the initial flow was) and possibly at the
site of the magnetar itself. In this scenario any radio emission
associated with more diffuse components would be due to 
leptons which have diffused away from the particle acceleration
site.

To conclude, we have observed at high angular resolution the early
stages of the evolution of the radio counterpart to the giant flare of
SGR 1806-20 on 2004 Dec 27. Despite many difficulties with analysing
and interpreting the MERLIN and VLBA datasets, we find evidence for
significant substructure on angular scales $\la 100$ mas, associated
with about 10\% of the total radio flux. The compact radio
structure is likely to be associated with sites of particle
acceleration, which may be external or internal shocks in the outflow,
or somehow associated with the environment close to the magnetar
itself.

\section*{Acknowledgements}

MERLIN is a National Facility operated by the University of Manchester
at Jodrell Bank Observatory on behalf of the UK Particle Physics and
Astronomy Research Council.  The US National Radio Astronomy
Observatory is a facility of the National Science Foundation operated
under cooperative agreement by Associated Universities, Inc. We thank
the referee for detailed and constructive criticism.

\end{document}